\documentclass[sigconf]{acmart}

\usepackage{acronym}
\usepackage[inline]{enumitem}
\usepackage{balance}
\usepackage{amsmath}
\usepackage{array,booktabs}
\usepackage{multirow}
\usepackage{adjustbox}
\usepackage{numprint}
\usepackage{xspace}
\usepackage[british]{babel}

 \AtBeginDocument{%
  \providecommand\BibTeX{{%
    \normalfont B\kern-0.5em{\scshape i\kern-0.25em b}\kern-0.8em\TeX}}}

\copyrightyear{2023}
\acmYear{2023}
\setcopyright{rightsretained}
\acmConference[SIGIR '23]{Proceedings of the 46th International ACM SIGIR Conference on Research and Development in Information Retrieval}{July 23--27, 2023}{Taipei, Taiwan}
\acmBooktitle{Proceedings of the 46th International ACM SIGIR Conference on Research and Development in Information Retrieval (SIGIR '23), July 23--27, 2023, Taipei, Taiwan}\acmDOI{10.1145/3539618.3591683}
\acmISBN{978-1-4503-9408-6/23/07}


\renewcommand{\shortauthors}{Anon et al.}

\acrodef{CIS}{conversational information seeking}
\acrodef{CS}{conversational search}
\acrodef{IR}{information retrieval}
\acrodef{NLP}{natural language processing}
\acrodef{PLM}{pretrained language model}
\acrodef{LLM}{large language model}

\newcommand{\ConvSim}[0]{\textit{ConvSim}\xspace}

\usepackage{amsmath}
\DeclareMathOperator*{\argmax}{arg\,max}

\author{Paul Owoicho*}
\affiliation{%
  \institution{University of Glasgow}
  \city{Glasgow}
  \country{Scotland, UK}
}
\email{p.owoicho.1@research.gla.ac.uk}

\author{Ivan Sekuli\'c*}
\affiliation{%
  \institution{Università della Svizzera italiana}
  \city{Lugano}
  \country{Switzerland}
}
\email{ivan.sekulic@usi.ch}

\author{Mohammad Aliannejadi}
\affiliation{%
  \institution{University of Amsterdam}
  \city{Amsterdam}
  \country{The Netherlands}
}
\email{m.aliannejadi@uva.nl}

\author{Jeffrey Dalton}
\affiliation{%
  \institution{University of Glasgow}
  \city{Glasgow}
  \country{Scotland, UK}
}
\email{jeff.dalton@glasgow.ac.uk}

\author{Fabio Crestani}
\affiliation{%
  \institution{Università della Svizzera italiana}
  \city{Lugano}
\country{Switzerland}
}
\email{fabio.crestani@usi.ch}


\begin{document}

\title{Exploiting Simulated User Feedback for Conversational Search: Ranking, Rewriting, and Beyond}

\begin{abstract}
This research aims to explore various methods for assessing user feedback in mixed-initiative \ac{CS} systems.
While \ac{CS} systems enjoy profuse advancements across multiple aspects, recent research fails to successfully incorporate feedback from the users.
One of the main reasons for that is the lack of system--user conversational interaction data.
To this end, we propose a user simulator-based framework for multi-turn interactions with a variety of mixed-initiative \ac{CS} systems.
Specifically, we develop a user simulator, dubbed \emph{\ConvSim}, that, once initialized with an information need description, is capable of providing feedback to system's responses, as well as answering potential clarifying questions.
Our experiments on a wide variety of state-of-the-art passage retrieval and neural re-ranking models show that effective utilization of user feedback can lead to $16\%$ retrieval performance increase in terms of nDCG@3.
Moreover, we observe consistent improvements as the number of feedback rounds increases ($35\%$ relative improvement in terms of nDCG@3 after three rounds).
This points to a research gap in the development of specific feedback processing modules and opens a potential for significant advancements in \ac{CS}.
To support further research in the topic, we release over \numprint{30000} transcripts of system-simulator interactions based on well-established \ac{CS} datasets.
\end{abstract}


\maketitle

\def\thefootnote{*}\footnotetext{These authors contributed equally to this work}\def\thefootnote{\arabic{footnote}}

\section{Introduction}

The primary goal of a conversational search (CS) system is to satisfy the user's information need. 
However, there are several challenges that arise when it comes to \ac{CS}, as opposed to traditional \emph{ad-hoc} search. 
An important tool for addressing these challenges is the use of mixed-initiative techniques. Under the mixed-initiative paradigm, the conversational search system can proactively initiate prompts, such as suggestions, warnings, or questions, at any point in the conversation. In recent years, mixed-initiative conversational search has received significant attention from the \ac{IR} research community, leading to advancements in various aspects of this field, including conversational passage retrieval~\cite{dalton2020trec, yu2021few}, query rewriting in context \cite{vakulenko2021comparison}, intent prediction in conversations~\cite{qu2019user}, and asking clarifying questions~\cite{aliannejadi2019asking}.

\begin{figure*}
    \centering
    \includegraphics[width=\textwidth]{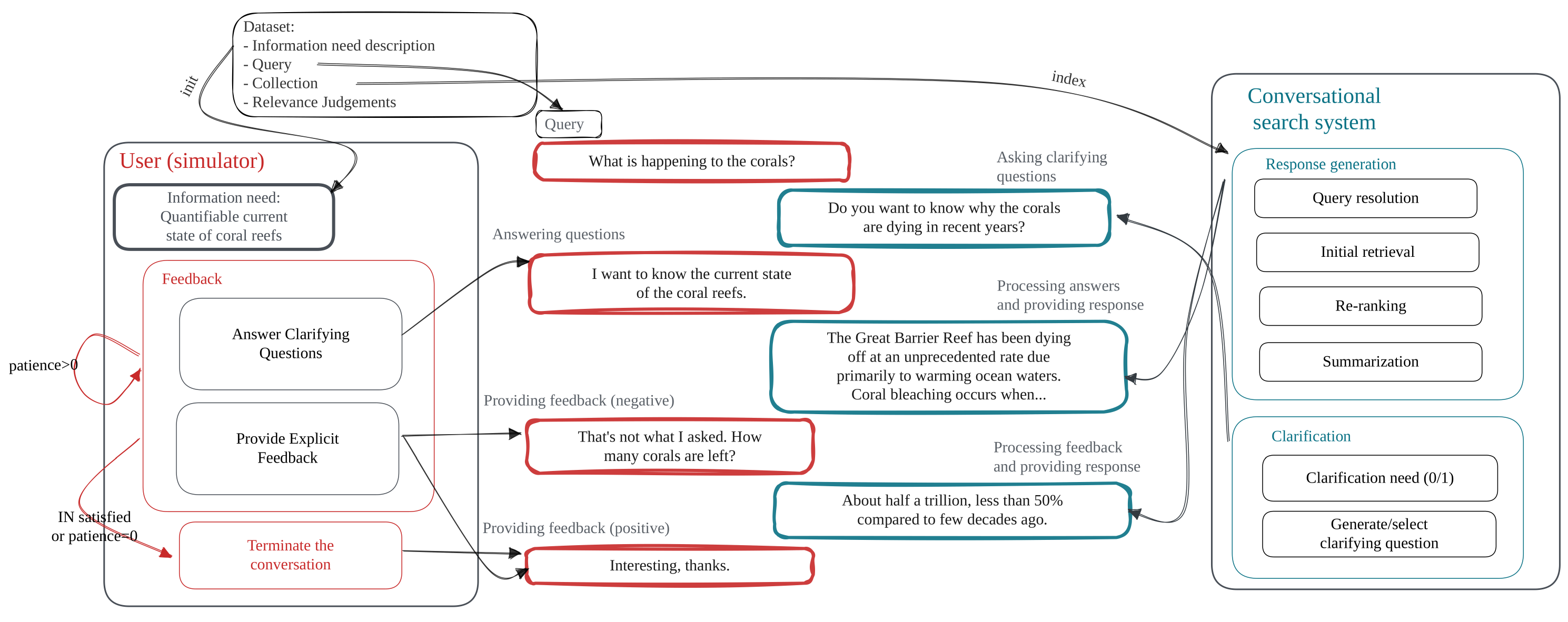}
    \vspace*{-6mm}
    \caption{Experimental framework with an example interaction between a user simulator (left) and a mixed-initiative conversational search system (right). Functionalities and modules of both are highlighted.}
    \label{fig:the_image}
    \vspace*{-1mm}
\end{figure*}

Despite the abundance of research on various components of mixed-initiative search systems, little has been done to study the impact of user feedback. 
Users can provide explicit feedback on the quality of system's responses, as well as answer potential questions prompted by the system.
Such feedback is beneficial to mixed-initiative \ac{CS} systems and can provide valuable information on user's needs.
Moreover, feedback can have a great effect on how conversation is shaped by, e.g., giving the system the chance to recover from an initial failed attempt~\cite{zou2023benefit}.
Despite its significance, lack of research in this area can be attributed to the difficulty of collecting appropriate data containing user feedback.

\looseness=-1
Furthermore, evaluation of \ac{CS} systems is arduous~\cite{penha2020challenges, lipani2021doing}. %
Typically, it requires the actual users to interact with the system, presenting their information needs, answering potential questions, and providing feedback.
Such studies are expensive and time consuming, often requiring a large number of experiments to properly evaluate specific approaches. 
That is even more the case with mixed initiatives, as the number of possible conversations is essentially limitless~\cite{balog2021conversational}.
An attempt to address this issue is to compile offline collections aimed at specific challenges in conversational search~\cite{dalton2020trec,aliannejadi2021building,qu2018analyzing}.
Existing data collections are mainly built based on online human--human conversations~\cite{qu2018analyzing}, synthetic human--computer interactions~\cite{dalton2020trec}, and multiple rounds of crowdsourcing~\cite{aliannejadi2021building}.
No existing data collections, however, feature explicit user feedback extensively in a conversation, thus limiting research in this area.
Moreover, such corpus-based evaluation paradigms usually remain limited to single-turn interactions and do not take into account the interactive nature of \ac{CS}, not to mention being limited to non-generative models.

To address the vicious circle composed of the lack of research on feedback utilization and the lack of appropriate data, we develop a comprehensive experimental framework based on simulated user--system interactions, as shown in Figure~\ref{fig:the_image}.
The framework allows us to evaluate multiple state-of-the-art mixed-initiative \ac{CS} systems, addressing several challenges, including contextual query resolution, asking clarifying questions, and incorporating user feedback.

Existing work~\cite{aliannejadi2021analysing} aims to study the effect of different mixed-initiative strategies on retrieval, however, their findings are limited to a single data collection, and lexical-based retrieval techniques. 
More recently, work on user simulators for conversational systems aims to address these limitations, however, it remains limited to pre-defined or templated interactions~\cite{zhang2020evaluating, salle2021studying} or focus only on one aspect of the search system, e.g., answering clarifying questions~\cite{sekulic2022evaluating}.
To address these limitations, we propose a user simulator called \emph{\ConvSim}, capable of multi-turn interactions with mixed-initiative \ac{CS} systems.
Given a textual description of the information need, \emph{\ConvSim} answers prompted clarifying questions and provides both positive and negative feedback, as necessary.
Recent advancements in \acp{LLM}, e.g., GPT-3~\cite{brown2020language}, PALM~\cite{chowderry2022palm}, open the possibilities of addressing such nuanced tasks.
Thus, we base core functionalities of the proposed simulator on \acp{LLM}.
Finally, the \emph{\ConvSim} addresses the limitation of pre-built corpora, as the simulator's behavior adapts to the  system's response. %

Our experimental evaluation shows that \emph{\ConvSim} can reliably be used for interacting with mixed-initiative conversational systems.
Specifically, we demonstrate that responses generated by the simulator are natural, in line with defined information needs, and, unlike previous work~\cite{sekulic2022evaluating}, coherent across multiple conversational turns.
The proposed simulator interacts with \ac{CS} systems entirely in natural language, without the need to access the system's source code or inner mechanisms.
Furthermore, the experimental framework, centered around \emph{\ConvSim}, allows for seamless curation of synthetic data on top of existing static IR benchmarks, as the simulator--system interactions can extend over multiple conversational turns.

We stress the fact that research questions around feedback utilization in \ac{CS} can hardly be answered by existing or pre-built collections.
On the other hand, while the questions around leveraging user feedback could be answered through comprehensive user studies, such studies are time-consuming, expensive, and largely limited in the number of experiments we would be able to conduct.

\looseness=-1
We find significant improvements in retrieval performance of methods utilizing feedback compared to non-feedback methods, even with only a single turn of feedback.
Well-established methods, such as RM3, adapted to handle explicit feedback, demonstrate relative improvement of $11\%$ and $9\%$ in terms of recall and nDCG@3.
Further, we identify a shortcoming of standard T5 query rewriter \cite{lin2020multi} in the task of processing feedback. %
To address this, we propose a novel adaptation of the T5 method and achieve $10\%$ and $16\%$ improvements in terms of recall and nDCG@3, respectively. %
Similarly, incorporating answers to clarifying questions yields improvements both in recall ($18\%$) and nDCG@3 ($12\%$). %
We also find that multiple rounds of simulator--system interactions result in further improvements in retrieval effectiveness ($35\%$ relative improvement in terms of nDCG@3 after three rounds).
Moreover, we observe that existing methods react poorly to certain types of feedback (e.g., positive feedback ``Thanks''), leading to a decrease in performance.
This points to a research gap in development of specific feedback processing modules and opens a potential for significant advancements in \ac{CS}.

Our main contributions are:
\begin{itemize}
    \item New insights into mixed-initiative \ac{CS} system design, with a focus on processing users' feedback, including explicit feedback and their answers to clarifying questions.
    \item A user simulator, capable of multi-turn interactions with mixed-initiative search systems. We release transcripts, code and guidelines\footnote{https://github.com/grill-lab/ConvSim} to foster further research.
\end{itemize}

\section{Related Work}
\label{sec:related_work}

\subsection{Mixed-initiative conversational search}
In recent years, conversational search has attracted significant attention both from the \ac{IR} and \ac{NLP} communities~\cite{anand2020conversational}.
To this end, \citet{radlinski2017theoretical} propose a theoretical framework of conversational search, identifying key properties of such systems and focusing on natural and efficient information access through conversations.
While some of the challenges remain similar to traditional \emph{ad hoc} search, significant new ones arise in the conversational paradigm. These are surveyed in the recent manuscript of \citet{zamani2022conversational}.
They include conversational query rewriting~\cite{yu2020few,vakulenko2021comparison}, conversational retrieval~\cite{dalton2020trec,yu2021few} and user intent prediction~\cite{qu2019user}. 

One key element of conversational search is mixed-initiative, which is the interaction pattern where both the system have rich forms of interaction. Under the mixed-initiative paradigm, conversational search systems can at any point of conversation take initiative and prompt the user with various questions or suggestions. Mixed-initiative has a long history in dialogue systems with \citet{walker1990mixed} identifying it as an integral part of conversations and \citet{horvitz1999principles} identifying key principles of mixed-initiative interactions. One of the most prominent uses of mixed-initiative is asking clarifying questions with a goal of elucidating the underlying user's information need~\cite{aliannejadi2019asking,braslavski2017you,stoyanchev2014towards,zamani2020mimics}.
The benefits of prompting the user with clarifying questions is found by multiple studies, including improving retrieval performance in conversational search~\cite{aliannejadi2020convai3,krasakis2020analysing,rosset2020leading,zamani2020analyzing,zou2020empirical}.
Clarifying questions are generally either selected from a pre-defined pool of questions~\cite{aliannejadi2019asking,aliannejadi2020convai3,rao2018learning} or generated~\cite{zamani2020generating,rao2019answer,sekulic2021towards}.  While decent success has been demonstrated by various question selection methods~\cite{aliannejadi2020convai3}, such approaches remain limited to pre-defined conversational trajectories and are not fit for a realistic search scenario.
Therefore, generating a clarifying question poses itself as a natural improvement over the selection task, mitigating the need to collect all of the potential questions beforehand.
Various question generation methods exist, centered around either template-based questions or \ac{LLM}-based generation.
In this work, we also study clarifying questions and use simulation methods to answer them. While there are benefits to clarifying questions, there is also cost to the user for these interactions \cite{azzopardi2011economics,azzopardi2022towards}. In this work we focus on their effectiveness in a simulation environment and don't study user costs directly.

\subsection{Evaluation and user simulation}
\citet{deriu2021survey} state that the evaluation method in context of conversational systems should be automated, repeatable, correlated to human judgments, able to differentiate between different conversational systems, and explainable.
However, evaluating all of these elements in conversational systems is challenging.
While various unsupervised and user-based evaluation methods exist~\cite{deriu2021survey} there are key trade-offs.
\citet{liu2016not} conduct a thorough empirical analysis of unsupervised metrics for conversational system evaluation and conclude that they correlate very weakly with human judgments, emphasizing that reliable automated metrics would accelerate research in conversational systems.
Thus, \citep{deriu2021survey} identify user studies as a more reliable method for evaluating conversational systems, stressing the fact that such evaluation is both cost- and time-intensive.

Conversational search has similar evaluation challenges, further complicated by the retrieval of relevant documents from a large collection~\cite{penha2020challenges}.
While traditional Cranfield paradigm fits well for evaluation of \emph{ad hoc} search systems, it is not easily transferable to conversational search~\cite{lipani2021doing,fu2022evaluating}.
One of the specific challenges is that the complexity of multi-turn queries and the overall context is ignored by traditional metrics, and requires a more holistic approach~\cite{hassan2010beyond,jarvelin2008discounted}.

~\citet{balog2021conversational} makes the case that simulation is an important emerging research frontier for conversational search evaluation. 
\citet{paakkonen2017validating} assess the validity of the use of simulated users in interactive \ac{IR} and find it justified under a common interaction model.
While user simulators are a well-established idea in \ac{IR}~\cite{mostafa2003simulation,carterette2011simulating}, including applications such as simulating user satisfaction for the evaluation of task-oriented dialogue systems~\cite{sun2021simulating} and recommender systems~\cite{zhang2020evaluating,afzali2023usersimcrs}, their utilization in mixed-initiative conversational search is limited.

To address this, \citet{salle2021studying} design a simulator that selects an answer to potential clarifying questions posed by the system.
However, their approach is limited to pre-defined clarifying questions and pre-defined answers, making its usability restricted to a closed collection of such questions and answers.
\citet{sekulic2022evaluating} address that issue and design \emph{USi}, a simulator capable of generating answers to clarifying questions posed by the system.
Nonetheless, their approach is limited to single-turn interactions and does not take into account conversational context.
Moreover, \emph{USi} only addresses clarifying questions that are direct and about a single facet of the query.
In this work we propose \emph{ConvSim}, a simulator capable of multi-turn interactions with mixed-initiative conversational search systems.
\emph{ConvSim} addresses the challenges of previous work, while also further extending simulator capabilities by being able to provide positive and negative feedback to system's responses.

\vspace{-2mm}
\section{Background and problem definition}
\label{sec:definition}

In this section, we formally define the main task definition of mixed-initiative conversational search systems.
We then link these to the requirements of user simulation.

Formally, a search session consists of multiple turns of the user's utterances $u$ and the system's utterances $s$, forming conversational history $H = [u^1, s^1, \dots, u^{t-1}, s^{t-1}]$, with $u^t$ and $s^t$ corresponding to user's and system's utterance at conversational turn $t$, respectively. One key factor is that we differentiate between discourse types of user utterances $u$, namely queries $u_q$, answers $u_a$ to clarifying questions posed by the system, and explicit feedback $u_f$ to the system's responses.
Similarly, the system's utterance $s$ can either be a response $s_r$ aimed at satisfying the user's information need $IN$ or a clarifying question $s_{cq}$ aimed at elucidating the user's information need. 
One of the inputs to various modules of mixed-initiative systems can as well be the ranked list of results $R = [r_1, r_2,\dots,r_N]$, retrieved in response to $u^t$, where $N$ is the maximum number of results considered.

\subsection{Mixed initiatives}
\label{sec:definition:mi}
A conversational search system should be able to effectively conduct contextual query understanding, document retrieval, and response generation.
Moreover, under the mixed-initiative paradigm, the \ac{CS} system can at any point take initiative and prompt the user with various suggestions or clarifying questions~\cite{radlinski2017theoretical}.

\subsubsection{Clarification}
\label{sec:definition:mi:clarification}
When necessary, e.g., in case of a user's query being ambiguous, the \ac{CS} system can ask a clarifying question, or questions, to elucidate the user's underlying information need.
Thus, the first challenge of a mixed-initiative search system is to assess the need for clarification~\cite{aliannejadi2021building}.
Specifically, given the current user's utterance $u^t$, the task is to predict whether asking a clarifying question is required, or whether the system should issue a response aimed at answering the user's question.
Thus one of the modules of the search system needs to model a function $clarification\_need = f(u^t | H, D)$, where $clarification\_need \in \{0, 1\}$, indicating whether not to ask or to ask a clarifying question.

As mentioned, asking clarifying questions methods can be broadly categorized into \textit{question selection} and \textit{question generation}~\cite{aliannejadi2019asking,aliannejadi2020convai3} methods.
In the first approach, given the current user utterance, $u^t$, and a conversational history $H$, the task is to select an appropriate clarifying question from a predefined pool of questions $CQ = \{cq_1, cq_2,\dots,cq_n\}$.
Formally, we model $s_{cq} = \phi(u_t | H, R, CQ)$ where $\phi$ is our question selection model.
As discussed in Section~\ref{sec:related_work}, question generation poses itself as a necessary step in \ac{CS}, going beyond selection from pre-defined corpora.
Formally, the task of the question generation module is to model $\psi$ in  $s_{cq} = \psi(u_t | H, R)$.
In this work, we implement several state-of-the-art question selection and generation models and evaluate their performance.
Moreover, we test the robustness of feedback processing modules depending on the type of clarifying question.

\subsubsection{Processing user feedback}
\label{sec:definition:mi:feedback}
A \ac{CS} system needs to be able to process feedback given by the user during the conversation including both answers to clarifying questions and explicit feedback to the system's response.
Therefore, the system, in both cases, needs to update its internal state by refining its representation of the user's information need.
Formally, we define updates to the system's interpretation of the user's information need, as query reformulation:
    $u^{t'} = \gamma(u^t | H)$, where $\gamma$ is the query rewriting model.
We note that, depending on the design choices of mixed-initiative systems, different forms of feedback, i.e., answers to clarifying questions and explicit feedback to the system's responses, can be modeled differently --- e.g., $u^{t'} = \gamma_1(u^t_a | H)$ and $u^{t'} = \gamma_2(u^t_f | H)$.
Furthermore, we point out that similar methods might be used to model contextual query reformulation, which aims at resolving current user utterance in the context of conversational history: $u^{t'}_q = \gamma_3(u^t_q | H)$.

\vspace{-2mm}
\subsection{User simulation}
\label{sec:definition:simulator}

A user simulator aims to mimic key user's roles in MI interactions.
Although \citet{balog2021conversational} defines several desired properties of a realistic user simulator, we focus on the simulator's ability to capture and communicate aspects of the information need. 
The simulator should coherently answer any posed clarifying questions, or provide positive/negative feedback to the system's responses.
In other words, the requirements of a user simulator are complementary to the ones of mixed-initiative \ac{CS} systems.
Inspired by \citet{zhang2020evaluating}, we base our user interaction model on the general QRFA model for the conversational information-seeking process~\cite{vakulenko2019qrfa}.

Formally, the user simulator needs to be able to carry out multi-turn interactions with the search system and generate a variety of different utterances: \begin{enumerate*}[label=(\roman*)]
    \item $u_q$ --- seek information through querying;
    \item $u_a$ --- answer clarifying questions; and
    \item $u_f$ --- provide feedback to systems' responses.
\end{enumerate*}
All of the utterances generated by the simulator need to be in line with the underlying information need $IN$.
First, a simulator needs to represent its information need by constructing a query utterance $u_q = h(IN)$.
Moreover, when prompted with a clarifying question utterance $s_{cq}$, the user simulator should be able to provide an answer $u_a = \theta_1(s_{cq} | H, IN)$, where $\theta_1$ denotes answer generation model.
Similarly, when given a response $s_r$ to its query, it needs to generate feedback $u_f = \theta_2(s_r | H, IN)$, where $\theta_2$ is the response generation function.
Figure~\ref{fig:the_image} shows a components of the simulator, where $\theta_1$ and $\theta_2$ are utilized at appropriate stages.

Asking too many clarifying questions or providing unsatisfactory responses might impair user's satisfaction with the search system~\cite{zou2023benefit}.
Thus, a simulator should encapsulate similar behaviors.
Following \citet{salle2021studying}, we introduce the notion of \emph{patience} $p \in \mathbb{Z}^{0+}$ --- a parameter that indicates how many turns of feedback a simulated user willing to provide.
Simulator decreases its patience $p$ after each turn in which it has to provide feedback, terminating the conversation once $p = 0$.
A conversation is stopped by the simulator either when $IN$ is satisfied or when patience runs out.

\subsubsection{Naturalness and usefulness of generated answers.}
\label{sec:evaluation:simulator:natural}
In order for simulator's behavior to be similar to real users~\cite{balog2021conversational}, both answers $s_a$ and feedback $s_f$ need to be relevant, in coherent natural language, and consistent with information need $IN$.
Following \citet{sekulic2022evaluating}, we assess \emph{naturalness} and \emph{usefulness} of the generated answers to clarifying questions.
\emph{Naturalness} refers to the utterance being in fluent natural language and likely generated by humans \cite{peng2020few, sai2022survey}.
We ground our definition of \emph{usefulness} in previous work assessing clarifying questions \cite{rosset2020leading} and their answers \cite{sekulic2022evaluating}.
Specifically, it captures whether answers and feedback generated by the simulator are consistent with the provided information need, and can be related to adequacy~\cite{stent2005evaluating} and informativeness~\cite{chuklin2019using}.
Moreover, by extending the evaluation to the multi-turn setting, we are also evaluating simulator's context awareness.

\subsubsection{Feedback.}
\label{sec:evaluation:simulator:feedback}
Explicit feedback $u_f$, generated in response to the systems' responses, needs to be reliable and accurate.
To this end, at each turn $u^t_q$, the system returns response $s_r^{t+1}$ and the simulator generates feedback $u_f^{t+1}$.
Moreover, the utterance $u_f^{t+1}$ is externally annotated as positive or negative feedback.
Our aim is to measure correlation of retrieval performance at turn $u_q^t$ and type of feedback $u_f^{t+1}$ (positive or negative).
Finally, we assess potential differences, as measured by retrieval metrics, between turns that received positive vs negative feedback.
Positive feedback should be generated in cases where performance is high, while negative feedback should be given when performance is low.

\section{Methodology}

\subsection{Proposed simulator framework}
\label{sec:methodology:mi-systems:conv-pipeline}

We propose \emph{\ConvSim}, a Conversational search Simulator, capable of multi-turn interactions with the search system in a conversational manner.
We design \ConvSim to satisfy the requirements defined in Section~\ref{sec:definition:simulator}.
As such, the simulator needs to encapsulate different behaviors across utterances of various discourse types, including querying $u_q$, as well as providing feedback $u_a$ and  $u_f$.

We conduct our simulator experiments within the framework of a conversational pipeline that encapsulates the commonly used components in a mixed-initiative conversational search pipeline: query rewriting, passage retrieval, passage reranking, clarifying question selection and generation, and response generation. 
The framework is depicted in Figure~\ref{fig:the_image}.
It enables seamless multi-turn exchange of user simulator utterances $u$ and system's utterances $s$, detailed in Section~\ref{sec:definition}.
The framework includes a suggested logical exchange of the utterances, i.e., when the system produces a response $s_r$, the simulator is tasked to provide feedback $u_f$.
Likewise, when posed with a clarifying question $s_{cq}$ the simulator needs to provide an answer $u_a$.
Such interactions continues as long as simulator patience $p > 0$ and $IN$ is not satisfied.
Moreover, we design this framework to be flexible, allowing us to easily configure and (re)arrange the steps per our experimental needs.
At the heart of this framework is a conversational turn representation that holds all relevant properties about a turn, such as a user query, system response, conversational context, and retrieved documents. 
We refer the reader to our codebase for the implementation details of this experimental framework.

Specifically, we initialize \emph{\ConvSim} with an information need description $IN_t$, specific to each turn.
This ensures the responses generated by \emph{\ConvSim} are consistent with the user information need and guide the conversation towards the relevant information.

We model feedback generation functions $\theta_1$ and $\theta_2$ detailed in Section~\ref{sec:definition:simulator} using \ac{LLM}s.
Given the focus of our experiments, we implement each of the simulator's possible actions (clarifying question answering for $\theta_1$, feedback generation for $\theta_2$) as steps in the conversational pipeline framework described below. 

\subsubsection{Implementation details}
We build \emph{\ConvSim} on top of OpenAI's Text-Davinci-003~\cite{brown2020language} model using few-shot prompting. 
We use OpenAI's completions API endpoint with the following parameter settings based on the author's guidelines~\cite{brown2020language} and initial empirical exploration:
\begin{itemize}[leftmargin=*]
    \item \textbf{\textit{max\_tokens}:} 50. This prevents the model from generating overly long responses but is also sufficient enough for the model to generate clarifying questions in addition to negative feedback or to expand a bit on its answers to clarifying questions.
    \item \textbf{\textit{temperature}:} 0.5. This is a halfway point between a very conservative and risky model. While we want creative outputs, we also want the responses to be on topic.
    \item \textbf{\textit{frequency\_penalty}:} 0.2. This discourages the model from generating previously generated tokens (i.e., repeating itself).
    \item \textbf{\textit{presence\_penalty}:} 0.5. This encourages the model to introduce new topics. In the same way as the \textit{temperature} parameter, this enables fairly novel responses that are always on topic.
\end{itemize}

For a given turn $t$, we prompt the model with a task description (i.e., whether to generate an answer to a clarifying question or feedback to system's response), a description of the information need $IN_t$, sample transcripts between a user and a system with the desired behavior, and a transcript of the conversational history $H$ between the user and system up to turn \textit{t}.  The exact prompts used can be found in our codebase.
We do not explicitly implement the information seeking model $u_q = h(IN)$.
Instead, we take the initial query $u_q^t$ directly from the dataset to ensure fair comparisons between non-feedback and feedback utilizing methods described above.

\subsection{Evaluation Data}
\label{sec:methodology:mi-systems:datasets}
We primarily use the \textbf{TREC CAsT}~\cite{owoicho2022trec} benchmark, designed for the development and evaluation of conversational search systems.
CAsT is composed of a series of fixed conversations, each with a pre-determined trajectory and containing a series of topical user utterances and canonical responses.  We focus on year 4 because it is the only dataset that includes mixed-initiative interactions.

Because each turn in CAsT does not have an $IN$ description, we augment it by adding turn-level information need descriptions.
Specifically, two expert annotators independently study each CAsT utterance in the conversation context and describe the full information need in a sentence. We decide on the length of the information, following the typical topic description in the TREC Web track topic list~\cite{clarke2009overview}. We instruct the annotators to take into account various sources of information such as the canonical responses and the rewritten queries. The final goal is to generate a self-contained description for each user utterance in CAsT. One could argue that the human rewritten utterances would be sufficient for this aim. In our preliminary analysis, we discover that the re-written utterances miss various contextual information that makes them dependent on the overall conversation context. %
We compare the generated information need descriptions by the two annotators. In case of minor differences, we select either of them. However, in cases where the difference is major there is discussion until agreement.

\vspace{-2mm}
\subsection{Mixed-initiative systems}
\subsubsection{Compared methods}
\label{sec:methodology:mi-systems:baselines} 
We focus our investigations on the effects and ways of using simulated user feedback and answers to clarifying questions for downstream retrieval. 
In order to analyze the effects of feedback processing modules, we compare their performances against the following non-feedback baselines which do not use any initiative or simulation:

\textbf{Organizer-auto} is a competitive baseline used in the TREC CAsT shared task over the past two years. 
First, it reformulates the user query with a generative T5 query rewriter fine-tuned on the CANARD dataset \footnote{https://huggingface.co/castorini/t5-base-canard}. As context, the rewriter takes in all previous turn queries and system responses as input: $u^{t'}_q = \gamma_3(u^t_q | H)$. 
No special considerations are made for cases where the input token length exceeds the model's limit (i.e., 512 tokens). Next, it uses Pyeserini's \footnote{https://github.com/castorini/pyserini} BM25 implementation (k1=4.46, b=0.82) to retrieve the top \numprint{1000} documents from the collection and re-ranks it's constituent passages with a point-wise T5 passage ranker (MonoT5)~\cite{nogueira2020document} trained on MSMARCO~\cite{bajaj2016ms}. 
Finally, a BART model \footnote{https://huggingface.co/facebook/bart-large-cnn} summarizes the top 3 passages to output a system response. 
We run \textbf{organizer-manual} on the CAsT benchmark using the manually reformulated queries at each turn for every conversation in the dataset. As these manual rewrites are context-free, this baseline represents an upper bound for retrieval performance without initiative or simulated responses using CAsT's bag-of-words retrieval and neural ranking methods. We refer the reader to CAsT'21 and CAsT'22 overview papers for more on the implementation details of these baselines.

For incorporating user feedback, we compare against additional baselines built on top of the \textbf{organizer-auto} baseline.
Formally, we model the following method with the function $u^{t'} = \gamma(u^t | H)$, described in Section~\ref{sec:definition:mi:feedback}, aimed at updating the system's understanding of the user's information need:
\begin{description}
\item[\textbf{organizer-auto+RM3}] uses the user feedback $u_f$ after the BART response generation step. 
Using the RM3 algorithm~\cite{lavrenko2009generative}, we expand the reformulated query $u^t$ with up to 10 terms from the feedback utterance $u_f$: $u^{t'}_q = u^t_q + RM3(u_f)$.
This expanded query is fed through the BM25 and MonoT5 steps, followed by BART response generation.
For our experiments, we interpret the number of feedback rounds as a proxy for user patience, detailed in Section~\ref{sec:definition:simulator}, i.e., the more rounds of feedback a user is willing to give, the more patient they are.

\item[\textbf{organizer-auto+Rocchio}] follows the same setup as \textbf{organizer-auto+RM3} but uses the Rocchio algorithm~\cite{rocchio1971relevance} for processing explicit feedback: $u^{t'}_q = u^t_q + Rocchio(u_f)$.

\item[\textbf{organizer-auto+QuReTeC}] expands the user's query with the QuReTeC model~\cite{Voskarides2020query} using terms from the conversation history. 
In our experiments, we adapt QuReTeC to additionally take terms from the explicit simulator feedback into account: $u^{t'}_q = u^t_q + QuReTeC(u_f, H)$. %
\end{description}

To assess if feedback utilization works on other systems, we also evaluate three of the strongest automatic submissions to CAsT'22, including \emph{splade\_t5mm\_ens}, \emph{uis\_sparseboat}, and \emph{UWCcano22}. We obtain the run files of these systems from the CAsT'22 organizers.

\subsubsection{Utilizing feedback} 
We implement query rewriting and passage ranking methods to utilize feedback by adapting state-of-the-art systems as follows:

\textbf{Passage Ranking.} 
We modify the query input of the MonoT5 re-ranker by adding feedback text to it, while keeping the passage input as is.
Specifically, we format the input to MonoT5 as follows:
\begin{verbatim}
    Query [u_q] [u_f] Passage [r_i] Relevant:
\end{verbatim}
where $u_q$, $u_f$, and $r_i$ refer to the query, feedback, and passage texts, respectively. 
Based on empirical investigations, we find this to be more effective in a zero-shot setting than changing the input template to accommodate feedback or using the feedback text in place of the query. 
We use an automatically rewritten query $u^{t'}_q$ as input, as opposed to the raw, unresolved query. Further, input lengths are restricted to 512 tokens. We refer to our variant of MonoT5-based model as \textit{FeedbackMonoT5}.

\textbf{Query rewriting.} We use the baseline T5 query rewriter (T5-CQR) to reformulate the feedback utterance based on conversation context (including the user`s raw query). We observe that this makes the rewriter prone to `over-rewriting', especially in the case of positive feedback. For example, `Thanks!' may be rewritten to `What types of essential oils should I consider for a scented lotion?', essentially repeating the user`s query, even after a positive feedback from the user. Given the lack of discourse-aware query rewriters, we examine the effects of mitigating this by also implementing an improved version of the rewriter that only reformulates negative feedback (Discourse-CQR). In both cases, as with the baseline system, the input text is automatically truncated where it exceeds the model's limit of 512 tokens.

Additionally, we process the answers to clarifying questions following \citet{aliannejadi2019asking}. Specifically, we append the answer and the asked clarifying question to the initial query: $u^{t'}_q = u^t_q + s^t_{cq} + u^t_a$.
The reformulated utterance is then $u^{t'}_q$ fed through our baseline pipeline \emph{organizer-auto}, without the first step of query rewriting.

\subsubsection{Asking clarifying questions}
\label{sec:methodology:mi-systems:CQs}
We implement several established approaches to asking clarifying questions.
While we acknowledge that not all utterances require clarification, as indicated by the $clarification\_need$ variable described in Section~ \ref{sec:definition:mi:clarification}, we do not explicitly model it.
The clarifying question is thus either not asked at all ($clarification\_need = 0$) or asked at each turn ($clarification\_need = 1$), depending on the experiment.
We focus on both question selection and question generation, implementing the following baselines.

\textbf{Question selection.} As detailed in Section~\ref{sec:definition:mi:clarification}, the aim of this group of models is to select an appropriate clarifying question utterance $s_{cq}^t$, given the user's current utterance $u^t_q$.
Therefore, we opt for two ranking-based methods.
First, a BM25-based method, termed \textbf{SelectCQ-BM25}, which indexes the clarifying question pool $CQ$ and performs retrieval with reformulated user utterance $u_q^{t'}$, specifically: $s_{cq}^t = arg\,max_i(BM25(cq_i| u_q^{t'})), cq_i \in CQ$.
A similar approach has been taken in previous works~\cite{aliannejadi2019asking, aliannejadi2020convai3}.
Second, a semantic matching-based method, termed \textbf{SelectCQ-MPNet}, utilizing MPNet~\cite{song2020mpnet} to predict a score for each question $cq_i$ from the pool: $s_{cq}^t = \argmax_i (MPNet(cq_i| u_q^{t'})), cq_i \in CQ$.
A similar approach has been adapted for CAsT'22~\cite{lajewska2022university}.
In both cases, the clarifying question with the highest score is selected, as indicated by the $arg\,max$ function.

\textbf{Question generation.}
We implement entity- and template-based clarifying question generation method, dubbed \textbf{GenerateCQ-Entity}.
Template-based question generation has been widely utilized in the research community due to its simplicity and effectiveness \cite{zamani2020generating,zhang2020evaluating,sekulic2021towards}.
With entities being central to the topic of a document, we opt to utilize SWAT~\cite{ponza2019swat} to extract salient entities to generate clarifying questions.
Specifically, we extract entities above a certain threshold ($\rho > 0.35$, as recommended by the authors) from the top $n$ results in the ranked list.
We then sort the entities by their saliency score in descending order, resulting in a list of entities $E = [e_1, e_2, \dots, e_M]$.
Finally, the question is constructed by inserting up to $m$ entities ($m$ is set to $3$) to the question template ``Are you interested in $e_1$, $e_2$, or $e_3$?''
Note that we alter the template according to the number of entities, in case $E$ contains less than 3 entities.

\vspace{-2mm}
\subsection{Evaluation}
\label{sec:evaluation}
\subsubsection{Mixed-initiative search systems}
\label{sec:evaluation:mi-systems}
We use the official measures and methodology from the CAsT benchmark for comparison. We report macro-averaged retrieval effectiveness of all systems at the turn level. 
We report NDCG@3 to focus on precision at the top ranks as well as standard IR evaluation measures (MAP, MRR, NDCG) to a depth of 1000 and at a relevance threshold of 2 for binary measures.
Statistical significance is reported under the two-tailed t-test with the null hypothesis of equal performance. We reject the null hypothesis in favor of the alternative with $p$-value $< 0.05$.
We design the experimental framework with the goal of assessing the impact of various \ac{CS} system components on retrieval performance.
Specifically, we evaluate the base pipeline, described in Section \ref{sec:methodology:mi-systems:conv-pipeline} for passage retrieval with and without \ac{CS} system components.

\vspace{-1mm}
\subsubsection{Naturalness and usefulness of generated answers.}
\label{sec:experimental:simulator:natural}
We evaluate \ConvSim in terms of naturalness and usefulness, as described in Section~\ref{sec:evaluation:simulator:natural}.
To this end, we compare our method to the current state-of-the-art simulator for answering clarifying questions, \emph{USi}~\cite{sekulic2022evaluating}, as well as human-generated responses.
Following \citep{sekulic2022evaluating}, we conduct a crowdsourcing-based evaluation on the ClariQ dataset~\cite{aliannejadi2020convai3}.
Specifically, two crowd workers annotate a pair of answers, where one is generated by \emph{\ConvSim}, and the other by \emph{USi} or humans. We instruct them to evaluate the answers in terms \emph{naturalness} and \emph{usefulness}.
In this pairwise setting, we count a win for a method if both crowd workers vote that the method's answer is more natural (or useful), while if the two crowd workers do not agree, we count it a tie.
For multi-turn evaluation, we utilize a multi-turn extension of the ClariQ dataset~\cite{sekulic2022evaluating} with human-generated multi-turn conversations. 
We follow \citet{li2019acute} and present full conversations for comparisons.
We report statistical significance under the trinomial test~\cite{bian2011trinomial}, an alternative to the binomial and Sign tests that takes into account ties.
The null hypothesis of equal performance is rejected in favor of the alternative with $p$-value $< 0.05$.
We present the results for both single- and multi-turn assessments.

We use the Amazon Mechanical Turk\footnote{\url{mturk.com}} platform for our crowd\-sourcing-based experiments.
We take several steps to ensure high-quality annotations:
\begin{enumerate*}[label=(\roman*)]
    \item we select workers based in the United States, in order to mitigate potential language barriers;
    \item the selected workers have above 95\% lifetime approval rate and at least \numprint{5000} approved HITs;
    \item we reject workers with wrong annotations on manually constructed test set;
    \item we provide fair compensation of \$0.25 per HIT, which with an average completion time of about 30 seconds, more than 300\% of the minimum wage in the U.S.
\end{enumerate*}

\vspace{-1mm}
\subsubsection{Feedback.}
We evaluate the feedback generation capabilities of \ConvSim as described in Section~\ref{sec:evaluation:simulator:feedback}.
To this end, we generate responses for each turn in the CAsT'22 dataset with the Organizer-auto method, described in Section \ref{sec:methodology:mi-systems:baselines}.
Next, we utilize \ConvSim to give feedback to the generated responses and manually annotate whether the generated feedback is positive or negative. 
We consider feedback positive if it is along the lines of \textit{``Thank you, that was helpful.''} and negative if similar to \textit{``That's not what I asked for.''} .
We consider it as negative feedback if it includes a more detailed sub-question aimed at eliciting the missing component (e.g., \textit{``Thanks, but what is its impact on climate change in developing countries?''}, since the information need is not entirely satisfied.
We compare the system's responses to the canonical responses present in CAsT to assess whether the information need is satisfied or not.

\section{Results}
\label{sec:results}
In this section we present the empirical evaluation with three core research questions:
\begin{enumerate}[label=\textbf{RQ\arabic*},leftmargin=*]
    \item How can we leverage user feedback and what is its effect on core components of a conversational search pipeline including: explicit relevance feedback processing, ranking and generating clarifying questions, and in core ranking?
    \item How does the \emph{\ConvSim} model compare with existing approaches for multi-turn simulation in terms of naturalness and usefulness?
    \item What is the effect of multiple rounds of simulated feedback when used in ranking?
\end{enumerate}

\subsection{Mixed-initiative systems}
Tables  \ref{tbl:query-reformulation-feedback} and ~\ref{tbl:passage-ranking-with-feedback} list the retrieval results for query reformulation and passage ranking, respectively. 
Generally, the results demonstrate improvements of feedback-aware methods over the baselines.
Below, we discuss the findings in detail.

\subsubsection{Query rewriting with feedback.} 

Compared to the baseline system, the addition of the \textit{QuReTeC} results in a ~39\% decrease in nDCG@3. 
This is surprising, considering \emph{QuReTeC}'s strong performance on previous editions of the CAsT benchmark. 
Likewise, \emph{Rocchio} also leads to a decrease in performance, with nDCG@3 going down by 0.151 points (~41\%).
In contrast, the addition of \emph{RM3} improves performance compared to the baseline, significantly outperforming it in terms of Recall, MAP, nDCG, and nDCG@3. 
Moreover, the results show the \emph{Discourse-CQR} method to outperform the baseline across all metrics, demonstrating the strongest performance among the implemented methods.

Expectedly, high-quality query rewriting/reformulation with feedback enables systems to retrieve more relevant passages in the initial retrieval stage at each turn, as evidenced by the increase in recall for the \emph{RM3} and \emph{Discourse-CQR} methods over the baseline. 
Not all reformulation methods are effective in all cases, however. 
Consider a turn where a user provides the following negative feedback without clarification: \emph{``That's not what I asked for. Can you please answer my question?''} Term expansion methods based on explicit feedback alone, such as \emph{RM3} and \textit{Rocchio}, completely fail, given the lack of relevant terms in the feedback utterance. 
On the other hand, methods that rely on explicit feedback and conversational history stand a better chance, as they have access to more relevant context to arrive at a better expression of the under-specified query. 

We note that, without fine-tuning, \emph{T5-CQR} performs competitively as a feedback rewriter, but still underperforms RM3 due to the `over-rewriting' issues discussed in Section~\ref{sec:methodology:mi-systems:baselines}. When we account for this with the \emph{Discourse-CQR} method, we observe boosts across all metrics. This suggests that naively using current models and systems to exploit explicit feedback through query rewriting are failure-prone. As a result, future `feedback-aware' conversational query rewriters need to take the feedback type into consideration, in order to be effective.

\begin{table}[]
\caption{Retrieval performance of methods for query reformulation using explicit feedback. Sign $\dagger$  indicates a significant difference compared to the \emph{organizer-auto} baseline.}
\vspace*{-4mm}
\label{tbl:query-reformulation-feedback}
\adjustbox{max width=\columnwidth}{%
\begin{tabular}{lrrrrr}
\toprule
Method                   & R & MAP & MRR & nDCG & nDCG@3 \\
\midrule
organizer-auto           & 0.348 & 0.155 & 0.533 & 0.311 & 0.365  \\
+ QuReTeC                & 0.192 & 0.088 & 0.310 & 0.180 & 0.223  \\
+ Rocchio                & 0.195 & 0.086 & 0.316 & 0.174 & 0.214 \\
+ T5-CQR                 & 0.340 & 0.131 & 0.500 & 0.288 & 0.329 \\
+ RM3                    & 0.388$\dagger$ & 0.167$\dagger$ & 0.565 & 0.343$\dagger$ & 0.398$\dagger$  \\
+ Discourse-CQR          & 0.384$\dagger$ & 0.174$\dagger$ & 0.620$\dagger$ & 0.348$\dagger$ & 0.423$\dagger$ \\

\bottomrule
\end{tabular}
}%
\end{table}

\subsubsection{Passage ranking with feedback.}
Across the board, we note that passage ranking with feedback leads to additional performance gains when used in a multi-step reranking setup. 
Specifically, the use of \emph{FeedbackMonoT5} on top of selected participant submissions to TREC CAsT'22 leads to boosts in nDCG@3, nDCG, and MRR scores at various reranking thresholds. 
Although we only report the results of ranking the top 100 passages in Table~\ref{tbl:passage-ranking-with-feedback}, we observe similar trends when reranking at depth 10 and 50, and expect that these observations continue beyond the depth of 100. 
We further note that the magnitude of the improvement explicit feedback brings for retrieval varies between these participant systems, indicating that the effectiveness of explicit feedback may depend on the underlying characteristics of each system.

We note that the addition of \emph{FeedbackMonoT5} leads to an average 6\% gain in nDCG@3. These results are consistent for the MRR metric too as \emph{FeedbackMonoT5} provides an average 7\% gain. Showing that explicit feedback can be useful in improving the overall retrieval. This is not just due to the quality of the MonoT5 passage ranker but is a result of the additional context from explicit feedback. 

We delve deeper into the queries where the delta in nDCG@3 before and after feedback ranking is at least 0.5 points in the \textit{splade\_t5mm\_ens} run. We observe that passage ranking with feedback hurts performance in cases of positive feedback (\textit{``Thanks,''} and negative feedback without clarification (\textit{``Can you please answer my question?''}), whereas negative feedback with clarification boosts performance (\textit{``That’s interesting, but what makes the beef so special?''}). Feedback that introduces more explicit context is more useful. As with query rewriting, this phenomenon suggests that ranking models should be feedback aware.

\begin{table}[]
\caption{Retrieval performance of passage ranking using explicit feedback on top of selected CAsT participant systems. This reranking step only reranks the first 100 passages from each system.}
\vspace*{-4mm}
\label{tbl:passage-ranking-with-feedback}
\adjustbox{max width=\columnwidth}{%
\begin{tabular}{lrrrrr}
\toprule
System                  & MAP & MRR & nDCG & nDCG@3 \\
\midrule
organizer-auto          & 0.155 & 0.533 & 0.311 & 0.365 \\
+ MonoT5                & 0.093 & 0.315 & 0.257 & 0.189 \\
+ FeedbackMonoT5        & 0.152 & 0.560 & 0.313 & 0.387  \\
\midrule
splade\_t5mm\_ens       & 0.217 & 0.585 & 0.479 & 0.411 \\
+ MonoT5                & 0.221 & 0.614 & 0.484 & 0.417  \\
+ FeedbackMonoT5        & 0.226 & 0.632 & 0.489 & 0.442 \\
\midrule
uis\_sparseboat         & 0.187 & 0.559 & 0.407 & 0.383 \\
+ MonoT5                & 0.177 & 0.581 & 0.399 & 0.381 \\
+ FeedbackMonoT5        & 0.184 & 0.611 & 0.408 & 0.415 \\
\midrule
UWCcano22               & 0.213 & 0.617 & 0.441 & 0.438 \\
+ MonoT5                & 0.217 & 0.612 & 0.443 & 0.427 \\
+ FeedbackMonoT5        & 0.217 & 0.659 & 0.454 & 0.454 \\
\bottomrule
\end{tabular}
}%
\end{table}

\subsubsection{Clarification and answer processing}
Table \ref{tbl:clarification} shows performance of three clarifying question construction methods, described in Section~\ref{sec:methodology:mi-systems:CQs}.
We observe an overall increase in effectiveness across all methods, with \emph{SelectCQ-BM25} and \emph{SelectCQ-MPNet} significantly outperforming the baseline across several metrics.
Most gains in performance are in recall, as the original query is expanded by the answer and clarifying question providing additional information to the initial retriever.
\emph{GenerateCQ-Entity} does not perform as well as selection-based methods. We attribute this finding to potentially off-topic clarifying questions, as the entities extracted were not necessarily geared towards elucidating user's need.
\emph{\ConvSim} might have responded along the lines of ``I don't know.'' or ``No thanks.'', thus not helping elucidate the underlying information need.

\begin{table}[]
\caption{Performance after asking a clarifying question constructed by various methods, compared to the baseline.}
\label{tbl:clarification}
\vspace*{-4mm}
\adjustbox{max width=\columnwidth}{%
\begin{tabular}{lrrrrr}
\toprule
Method              & R & MAP & MRR & nDCG & nDCG@3 \\
\midrule
organizer-auto  & 0.348 & 0.154  & 0.532 & 0.311 & 0.365  \\
+ SelectCQ-BM25        & 0.433$\dagger$ & 0.166 & 0.625 & 0.364$\dagger$ & 0.411 \\
+ SelectCQ-MPNet       & 0.413$\dagger$ & 0.173$\dagger$ & 0.631 & 0.362 & 0.409 \\
+ GenerateCQ-Entity & 0.409 & 0.162 & 0.577 & 0.348 & 0.398 \\
\bottomrule
\end{tabular}
}%
\end{table}

\subsection{User simulator}
\label{sec:results:simulator}

\subsubsection{Single- and multi-turn clarifying question answering.}
\label{sec:results:simulator:natural}
Table \ref{tbl:useful_natural} presents the results in comparison to \emph{USi}~\cite{sekulic2022evaluating} and human-generated answers to clarifying questions in single- and multi-turn scenarios.
We make several observations from the results.
First, \emph{\ConvSim} significantly outperforms \emph{USi} both in terms of naturalness and usefulness in both single- and multi-turn settings. 
Second, the difference between the performance of \emph{\ConvSim} and \emph{USi} is especially evident in the multi-turn setting, which is one of \emph{USi}'s potential limitations indicated by the authors~\cite{sekulic2022evaluating}. The difference is even greater in multi-turn usefulness assessments, which can be attributed to \emph{USi}'s hallucinations, and thus not staying on topic.
Finally, \emph{\ConvSim} in most cases does not significantly outperform human-generated answers, except in single-turn usefulness. Although further analysis is required, we suspect the difference to have come from \emph{\ConvSim}'s precision in answering clarifying questions, while crowd workers sometimes answer them reluctantly and concisely, with no notion of grammar and punctuality (e.g., ``no'').
\noindent The results indicate that \emph{\ConvSim} can be used to answer clarifying questions both in single- and multi-turn settings, outperforming state-of-the-art methods both in terms of naturalness and usefulness.

\begin{table}[]
\caption{Results of crowdsourcing study assessing naturalness and usefulness of generated answers to clarifying questions in single- and multi-turn scenarios. 
Each value indicates the percentage of pairwise comparisons won by the specific model as well as ties.
Sign $\dagger$ indicates a significant difference.}
\vspace*{-4mm}
\label{tbl:useful_natural}
\adjustbox{max width=\columnwidth}{
\begin{tabular}{llccc|ccc}
\toprule
&             & \ConvSim & \emph{USi}~\cite{sekulic2022evaluating} & Ties & \ConvSim & Human & Ties \\
\midrule
\multirow{2}{*}{\rotatebox[origin=c]{90}{\textit{Single}}} & Naturalness &             37\%$\dagger$  & 22\%  & 41\% & 36\% & 25\% & 39\%  \\
& Usefulness  & 44\%$\dagger$ & 19\% & 37\%  & 36\%$\dagger$ & 20\% & 44\%  \\
\midrule
\multirow{2}{*}{\rotatebox[origin=c]{90}{\textit{Multi}}}  & Naturalness & 45\%$\dagger$ & 18\% & 37\% & 25\% & 28\% & 47\% \\
& Usefulness  & 62\%$\dagger$ & 12\% & 26\% & 26\% & 16\% & 58\%     
\\ \bottomrule
\end{tabular}
}
\end{table}

\subsubsection{Generated feedback evaluation}
\label{sec:results:simulator:feedback}
Table \ref{tbl:feedback_results} shows the performances of \emph{Organizer-auto} model on CAsT'22 queries broken down by whether feedback given to the system's response is positive or negative, as described in Section~\ref{sec:evaluation:simulator:feedback}.
Results show significant differences between responses with positive and negative feedback. 
Feedback on the system's responses generated by \emph{\ConvSim} is useful, as the responses receiving negative feedback correspond to the poor retrieval effectiveness.
On the contrary, when the system's response satisfies the given information need, as demonstrated by higher retrieval performance, the simulator's feedback is positive.
\emph{\ConvSim} is not aware of the system's retrieval effectiveness and provides feedback solely on the generated response and $IN$ description.

\begin{table}[]
\caption{Performance on turns where feedback is negative vs.~turns where feedback is positive. The ``Perc.'' column indicates the percentage of such turns in the CAsT'22 dataset. All the differences are significant.}
\label{tbl:feedback_results}
\vspace*{-4mm}
\adjustbox{max width=0.9\columnwidth}{%
\begin{tabular}{lrrrrrr}
\toprule
Feedback & Perc. & R    & MAP  & MRR  & nDCG & nDCG@3 \\
\midrule
Negative & 49\%  & 0.073 & 0.039 & 0.399 & 0.091 & 0.161   \\
Positive & 51\%  & 0.185 & 0.128 & 0.739 & 0.239 & 0.449  \\
\bottomrule
\end{tabular}
}
\end{table}

\section{Discussion and Analysis}

\textbf{Does feedback help where it matters?}
Section \ref{sec:results} shows that systems that leverage feedback outperform systems that do not use it. We investigate a subset of 24 queries that require initiative as annotated by organizers~\cite{owoicho2022trec}. 
These turns require additional user input and are typically open-ended or a branching point. Systems that exploit user input should perform better on these queries than systems that do not. Table \ref{tbl:mi-turns-passage-ranking-with-feedback} shows results of feedback passage ranking method on top of the participant runs introduced in table \ref{tbl:mi-turns-passage-ranking-with-feedback}. Using feedback ranking \emph{FeedbackMonoT5} leads to non-significant improvements across most metrics for all runs with an average increase of 7.75\% in nDCG@3 with other metrics being similar.

\begin{table}[]
\caption{Passage ranking using explicit feedback on top of select CAsT participant runs. Runs are evaluated on a subset of queries annotated to require initiative.}
\label{tbl:mi-turns-passage-ranking-with-feedback}
\vspace*{-4mm}
\adjustbox{max width=\columnwidth}{%
\begin{tabular}{lrrrrr}
\toprule
System                  & MAP & MRR & nDCG & nDCG@3 \\
\midrule
organizer-auto          & 0.091 & 0.567 & 0.251 & 0.392 \\
+ FeedbackMonoT5        & 0.101 & 0.589 & 0.256 & 0.404 \\
\midrule
splade\_t5mm\_ens       & 0.168 & 0.597 & 0.444 & 0.424 \\
+ FeedbackMonoT5        & 0.195 & 0.659 & 0.463 & 0.492 \\
\midrule
uis\_sparseboat         & 0.137 & 0.739 & 0.381 & 0.445 \\
+ FeedbackMonoT5        & 0.133 & 0.733 & 0.378	& 0.490 \\
\midrule
UWCcano22               & 0.145	& 0.661 & 0.388 & 0.386 \\
+ FeedbackMonoT5        & 0.150	& 0.610	& 0.395 & 0.393 \\
\bottomrule
\end{tabular}
}%
\end{table}

\begin{figure}
\centering
\includegraphics[width=0.41\textwidth]{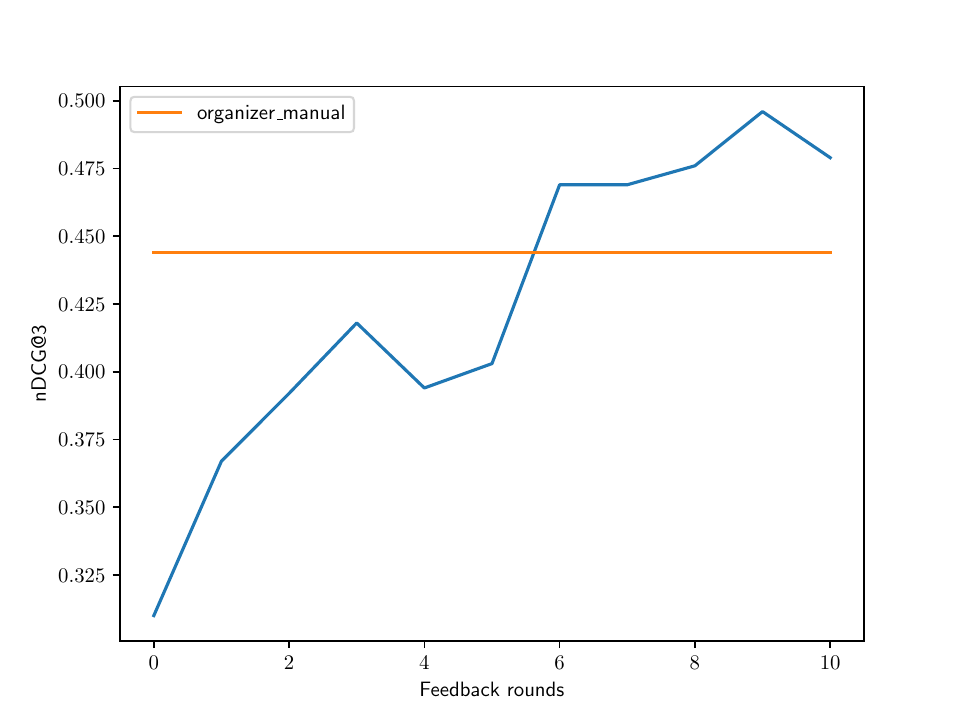}
\vspace*{-4mm}
\caption {Multiple rounds of feedback using the \textit{organiser-auto+Discourse-CQR+FeedbackMonoT5} system. The orange line depicts the performance of \textit{organizer\_manual}.}
\vspace*{-2mm}
\label{fig:feedback-rounds}
\end{figure}

\textbf{Effect of iterative feedback.}
We investigate the potential for multiple rounds of feedback in a simulated environment. 
We run the \emph{organiser-auto+Discourse-CQR} system with \emph{ FeedbackMonoT5} passage ranker for 10 rounds of feedback. 
For efficiency we only apply re-ranking to the first 100 passages retrieved. 
Figure~\ref{fig:feedback-rounds} shows consistent improvements in terms nDCG@3 over the \textit{organizer-auto} (round 0) baseline, with slight dips and plateaus between rounds 3 to 5 and rounds 6 to 8. At rounds 6 and above both MRR and nDCG@3 of this system exceed those of the \textit{organizer-manual} system. Recall and MAP at round 8 come within 0.004 and 0.003 points of the manual run, respectively, further highlighting the utility of explicit feedback.
Prompting the user for up to 8 or more rounds of feedback is not realistic and motivates the need for more effective feedback models that can learn from fewer rounds of feedback. 

\textbf{Combining clarification and explicit feedback.}
We analyze the effectiveness of \emph{FeedbackMonoT5} for processing answers to questions selected with \emph{SelectCQ-BM25}.
The results suggest an improvement over the \emph{organizer-auto} baseline (nDCG@3 $= 0.392$; $+7\%$ relative improvement), suggesting that \emph{FeedbackMonoT5} can be used for processing answers to clarifying questions.
We experiment with a round of clarification and a round of feedback and observe significant boost in Recall ($0.448$; $+29\%$ vs the baseline), but a relatively low improvement in terms of nDCG@3 ($0.389$; $+6\%$).
We hypothesize that both rounds of feedback result in well-defined information need, thus boosting the Recall, but query reformulation methods (i.e., \emph{FeedbackMonoT5}) fail to resolve the complex context, leading to poor re-ranking performance.

\section{Conclusions}
We study the effectiveness of mixed-initiative conversational search models in combination with simulated user feedback. 
Specifically, we compare and extend proven models with an aim of incorporating user feedback, including answers to clarifying questions and explicit feedback on system's responses.
We propose a new user simulator, \ConvSim, capable of multi-turn interaction, leveraging \ac{LLM}s.
The results show utilizing feedback consistently improves retrieval across the majority of the methods, resulting in $+16\%$ improvement in nDCG@3 after a single turn of feedback.
Moreover, we show that several rounds of feedback result in even greater boost ($+35\%$ after three rounds).
This promises potential for advancements in \ac{CS} and calls for further work on feedback processing methods.

\vspace{-2mm}
\section{Acknowledgments}
This work is supported by the Engineering and Physical Sciences Research Council (EPSRC) grant EP/V025708/1 and a 2019 Google Research Award.

\bibliographystyle{ACM-Reference-Format}
\bibliography{mybib}

\end{document}